\documentclass{emulateapj}
\usepackage{ulem}
 
\newcommand{\cf}{{\ifmmode{C_f}\else{$C_{f}$}\fi}}
\newcommand{\zem}{{\ifmmode{z_{em}}\else{$z_{em}$}\fi}}
\newcommand{\zabs}{{\ifmmode{z_{abs}}\else{$z_{abs}$}\fi}}
\newcommand{\kms}{{\ifmmode{{\rm km~s}^{-1}}\else{km~s$^{-1}$}\fi}}
\newcommand{\delv}{{\ifmmode{\Delta v}\else{$\Delta v$}\fi}}
\newcommand{\cm}{{\ifmmode{{\rm cm}^{-1}}\else{cm$^{-1}$}\fi}}
\newcommand{\cmm}{{\ifmmode{{\rm cm}^{-2}}\else{cm$^{-2}$}\fi}}
\newcommand{\cmmm}{{\ifmmode{{\rm cm}^{-3}}\else{cm$^{-3}$}\fi}}

\newcounter{species} 
\def\ion#1#2{\setcounter{species}{#2}#1$\;${\scriptsize\Roman{species}}\relax}

\slugcomment{To appear in the {\it Astrophysical Journal}, vol.\ 719, August 2010} 
\shorttitle{Spectropolarimetric Test of Quasar HS~1603+3820}
\shortauthors{Misawa et al.}

\begin{document}

\title{A Spectropolarimetric Test of the Structure of the Intrinsic
   Absorbers in the Quasar HS~1603+3820\altaffilmark{1}}

\footnotetext[1]{Based on data collected at Subaru Telescope, which is
operated by the National Astronomical Observatory of Japan.}

\author{Toru Misawa\altaffilmark{2,3,4},
        Koji S. Kawabata\altaffilmark{5},
        Michael Eracleous\altaffilmark{3,6},
        Jane C. Charlton\altaffilmark{3}, and
        Nobunari Kashikawa\altaffilmark{7}}

\altaffiltext{2}{Cosmic Radiation Laboratory, RIKEN, 2-1 Hirosawa,
  Wako, Saitama 351-0198 Japan}
\altaffiltext{3}{Department of Astronomy and Astrophysics, The
  Pennsylvania State University, University Park, PA 16802}
\altaffiltext{4}{School of General Education, Shinshu University,
  3-1-1 Asahi, Matsumoto, Nagano 390-8621 Japan}
\altaffiltext{5}{Hiroshima Astrophysical Science Center, Hiroshima
  University, 1-3-1 Kagamiyama, Higashi-Hiroshima, Hiroshima 739-8526
  Japan}
\altaffiltext{6}{Center for Gravitational Wave Physics, The
  Pennsylvania State University, University Park, PA 16802}  
\altaffiltext{7}{National Astronomical Observatory, Mitaka, Tokyo
 181-8588 Japan}

\email{misawatr@shinshu-u.ac.jp,
  kawabtkj@hiroshima-u.ac.jp,
  mce@astro.psu.edu,
  charlton@astro.psu.edu,
  kashik@zone.mtk.nao.ac.jp}

\begin{abstract}
We report the results of a spectropolarimetric observation of the
\ion{C}{4} ``mini-broad'' absorption line (mini-BAL) in the quasar
HS~1603+3820 ($z_{em} = 2.542$). The observations were carried out
with the FOCAS instrument on the Subaru telescope and yielded an
extremely high polarization sensitivity of $\delta p \sim$ 0.1\%, at a
resolving power of $R \sim 1500$. HS~1603+3820 has been the target of
a high-resolution spectroscopic monitoring campaign for more than four
years, aimed at studying its highly variable \ion{C}{4} mini-BAL
profile. Using the monitoring observations, in an earlier paper we
were able to narrow down the causes of the variability to the
following two scenarios: (1) scattering material of variable optical
depth redirecting photons around the absorber, and (2) a variable,
highly-ionized screen between the continuum source and the absorber
which modulates the UV continuum incident on the absorber. The
observations presented here provide a crucial test of the scattering
scenario and lead us to disfavor it because (a) the polarization level
is very small ($p$ $\sim$ 0.6\%) throughout the spectrum, and (b) the
polarization level does not increase across the mini-BAL trough.
Thus, the variable screen scenario emerges as our favored explanation
of the \ion{C}{4} mini-BAL variability. Our conclusion is bolstered by
recent X-ray observations of nearby mini-BAL quasars, which show a
rapidly variable soft X-ray continuum that appears to be the result of
transmission through an ionized absorber of variable ionization
parameter and optical depth.
\end{abstract}

\keywords{Galaxies: Quasars: Absorption Lines, Galaxies: Quasars:
Individual (HS 1603+3820)}

\section{Introduction}

Quasars are routinely used as background sources, allowing us to study
intervening gaseous structures via absorption-line diagnostics.  The
observed absorption lines have their origin in both {\it intervening}
objects, such as foreground galaxies, the inter-galactic medium (IGM),
or the interstellar medium (ISM) of the quasar host galaxies, and in
structures that are {\it intrinsic} to the quasars.

A promising candidate for the intrinsic absorbers is an outflowing
wind from the accretion disk that makes up the quasar central engine.
The outflow is thought to be accelerated by either magnetocentrifugal
forces \citep[e.g.,][]{eve05,dek95}, by radiation pressure in lines
and continuum \citep{mur95,pro00}, and/or by radiation pressure acting
on dust \citep[e.g.,][]{voi93,kon94}. According to many of the above
models, the flow can be equatorial, although the opening angle
relative to the plane of the disk depends on the choice of model
parameters. In other models, the flow can be polar \citep[directed at
right angles to the disk plane; e.g.,][]{kon94,pun99}.

The outflowing winds are important in two ways.  First, they may be
essential components of the quasar central engine: they can carry
angular momentum away from the disk, allowing accretion onto the black
hole to proceed. Second, they can deliver energy and momentum to the
ISM and IGM, thus regulating gas infall and star formation during the
assembly of a galaxy (e.g., \citealt{gra04, sca04, spr05}).  Thus,
feedback from quasar outflows can profoundly affect the properties of
present-day galaxies and it can also enrich the IGM with metals. For
these reasons, it is important to understand the properties and
structure of quasar outflows, a goal which can be accomplished by
studying the absorption lines that they produce.

\subsection{BALs, NALs, and Mini-BALs}

spectra of quasars, the {\it broad} absorption lines (hereafter, {\it
  BALs}; FWHM $\geq$ 2000~\kms) are easily identified as
intrinsic. The large widths of BALs make them very unlikely to
originate in intervening objects or the interstellar medium of the
quasar host galaxy.  BALs are detected in 10--20\% of optically
selected quasars \citep[e.g.,][]{hew03, rei03a}, and the detection
rate is slightly higher in radio-quiet quasars \citep[e.g.,][]{sto92,
  bec01, gre06}.\footnote{\citet{cha00} found that approximately 35\%
  of radio-quiet gravitational lensed quasars contain BAL features,
  which suggests that flux-limited optical surveys may be
  under-estimating the fraction of BALs.}  The observed fraction of
quasars hosting BALs constrains the solid angle subtended by the dense
portion of the wind to the central engine.

narrow absorption lines observed in the spectra of quasars (hereafter
{\it NALs}; FWHM $\leq$ 500~\kms) are also physically associated with
the quasars. This was motivated by the fact that the detection rate of
NALs varies between sight-lines and is related to the physical
properties of the background quasars \citep{ric99,ric01}. Such a trend
is not expected, if all NALs arise in intervening absorbers.  This
suggestion was confirmed once NALs were completely de-blended into
multiple narrower components with high-resolution spectroscopy.  We
can now separate intrinsic NALs from intervening ones, via two
characteristic properties related to the compactness of the gas
parcels in intrinsic absorbers: (a) dilution of absorption troughs by
un-occulted light from the background source(s) associated with the
quasar (hereafter ``partial coverage''; e.g., \citealt{bar97, ham97b,
  gan99}), and (b) time variability of line profiles (e.g., depth,
equivalent width, line centroid), within $\lesssim$1~yr in the quasar
rest frame (e.g., \citealt{ham97a, nar04, wis04}).  Neither of these
properties is expected in intervening absorbers because they have much
larger sizes and lower densities compared to intrinsic
absorbers. Using the former test, \citet{mis07a} find that $\sim$50\%\
of non-BAL QSOs at $z=2$--4 have at least one intrinsic NAL in their
rest-frame, near-UV spectra.

Equally useful, but less common, are {\it mini-BALs}, with intermediate
line widths between NALs and BALs (500~\kms\ $\leq$ FWHM $\leq$
2,000~\kms).  Mini-BALs are also identified by partial coverage and/or
time variability. However, they can also be identified just from their
smooth and relatively broad profiles, which is incompatible with the
properties of intervening gaseous structures.  Mini-BALs have the
advantages of both BALs (i.e., high probability of being intrinsic)
and NALs (i.e., line profiles can be resolved into individual
components), which makes them useful as diagnostics (e.g.,
\citealt{chu99, ham97a, nar04}).  

BALs, mini-BALs, and NALs could be complementary probes of quasar
outflows since they may be observed along different lines of sight
towards the quasar continuum and/or emission line source.  This idea
is bolstered by the results of X-ray observations by \citet{mis08} and
\citet{cha09}, which show that the NAL velocities and the column
densities of the corresponding X-ray absorbers follow a different
relation than BALs.

\subsection{Pilot Variability Study of HS~1603+3820}

As a pilot variability study of mini-BALs, we have been monitoring a
\ion{C}{4} mini-BAL in the optically bright quasar HS~1603+3820
($z_{em} = 2.542;\; B=15.9$) using primarily the High-Dispersion
Spectrograph (HDS) on the Subaru telescope \citep{mis03, mis05,
  mis07b}.  This mini-BAL spans a range of blueshifts of
8,300--10,600~\kms\ relative to the quasar and shows both partial
coverage and time variability.  Our monitoring campaign spans more
than four years (corresponding to about one year in the quasar
rest-frame).  No correlations are seen between the profile model
parameters, except for a possible correlation between total equivalent
width and the coverage fraction \citep[see Figure~5 of][]{mis07b}.  As
argued by \citet{mis07b}, simple changes of ionization state or motion
of a homogeneous gas parcel cannot reproduce the observed variations.
The former requires much faster variations of the UV continuum flux
than one would expect from such a luminous quasar (e.g.,
\citealt{giv99, haw01, kas07}). The latter requires that all the
absorbing parcels of gas (that correspond to each of the absorption
troughs) are at the same distance from the continuum source and cross
our cylinder of sight at the same time and without discernible changes
in their line-of-sight velocities, which is rather unlikely.  Only two
plausible scenarios remain for the origin of the variability: (a)
continuum photons, redirected by a variable scattering medium into our
cylinder of sight, dilute the mini-BAL absorption and modulate its
equivalent width according to the (variable) intensity of the
scattered light (depicted in Figure~1A), and (b) rapid continuum
fluctuations, coupled with coverage fraction fluctuations caused by a
porous/clumpy screen of variable optical depth located between the
continuum source and the mini-BAL gas (Figure~1B). A detailed
discussion of these scenarios is presented in \citet{mis07b}.  A third
possibility, not discussed explicitly by \citet{mis07b}, is that the
scattering medium surrounds the absorber itself, as depicted in
Figure~1C.

\begin{figure}
 \begin{center}
  \includegraphics[width=8cm,angle=0]{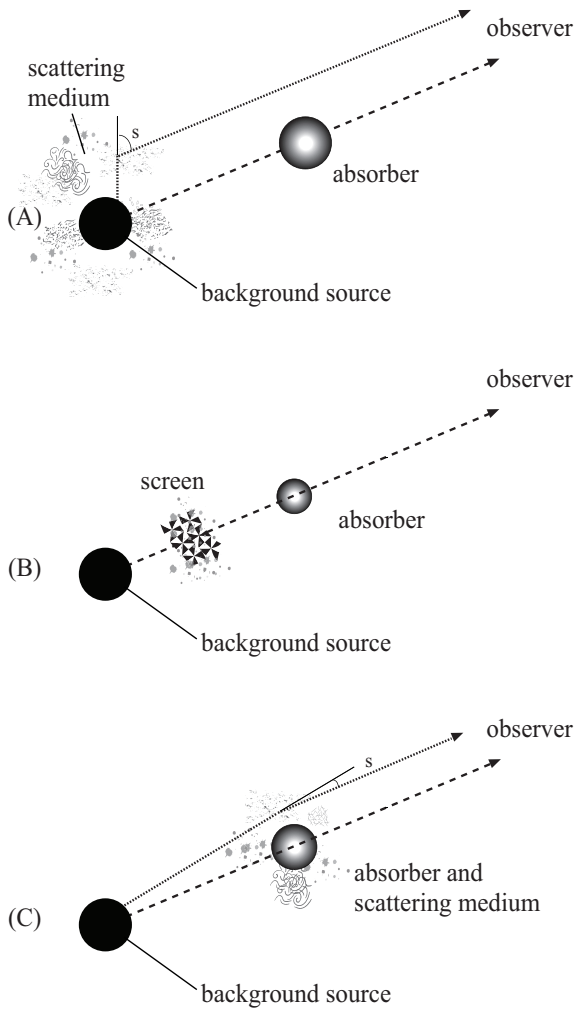}
 \end{center}
%
 \caption{Three possible geometries that could lead to apparent
   partial coverage and variability as seen in the \ion{C}{4} mini-BAL
   in HS~1603+3820.  The dashed lines represent the direct lines of
   sight towards the continuum source at the center of the accretion
   disk, while the dotted lines represent the paths of scattered
   photons.  (A) The absorbing parcel of gas is larger than the
   background continuum source but photons are redirected around it
   via scattering in a corona surrounding the central continuum
   source.  (B) The absorbing parcel of gas is small enough that it
   only occults a portion of the background continuum source.  A
   variable, porous/clumpy shield (e.g., the warm absorber) that
   modulates the transmitted flux is located between the \ion{C}{4}
   absorber and the continuum source. (C) Similar model to (A), but
   the scattering medium is surrounding the absorber rather than the
   continuum region. The scattering media in (A) and (C) are neither
   homogenous nor spherically symmetric about the continuum source or
   the absorber. The polarization degree depends on the scattering
   angle (denoted by $s$).}
\end{figure}

\subsection{Spectropolarimetric Test of the Scattering Hypothesis}

The scattering scenario for the variability of the \ion{C}{4} mini-BAL
in HS~1603+3820 can be tested through spectropolarimetric
observations.  Polarimetry and spectropolarimetry of BAL and non-BAL
QSOs shows that the former are {\it on average} more polarized than
the latter, in the continuum. However, the distribution of BAL
polarization levels is broad, with values between 0\% and 8\% (or even
higher in some cases). In contrast, non-BAL QSOs have polarization
levels less than 2.5\% \citep[e.g.,][see, for example, Figure1 in the
latter paper]{ogl99, sch99}. Of more interest here is the change of
polarization level between the continuum and the BAL troughs. The
implications of those observations form the basis for our
observational test. The BAL troughs are generally significantly more
polarized than the continuum. The degree of polarization in the BAL
troughs can reach 12--15\% \citep[e.g.][]{hin95, coh95, bro97, ogl99,
  sch99, sco04}. There are however examples where the BAL troughs are
actually unpolarized \citep[see][for a particularly interesting
case]{bro97}.

Two families of explanations for the observed increase in the
polarization level in BAL troughs have been discussed in the
literature: (a) Large-angle scattering from a medium that is off the
cylinder of sight to the background source. Thus the scattered
photons, which are also unattenuated, make up a significant fraction
of the residual light in the BAL troughs and increase the degree of
polarization (e.g., \citealt{goo95, lam04}). (b) Resonant scattering
within the absorber itself can also produce polarization with a level
as high as 15\% in the trough \citep[e.g.,]{lee97, wan07}.  The former
scenario appears more promising, because of the observed
anti-correlations the degree of polarization ($p$) and the detachment
index (DI)\footnote{The detachment index (DI) is the onset velocity of
  a BAL trough in units of the half-width of the corresponding broad
  emission line \citep{wey91}.}, and between the spectropolarimetric
index (SI)\footnote{The spectropolarimetric index (SI) is a measure of
  the strength of the absorption in the polarized flux, $f_{\lambda}
  \times p_{\lambda}$ relative to the absorption in the direct flux
  $f_{\lambda}$ \citep{lam04}.} and DI, which argue for the presence
of a scattering medium in the polar direction \citep{hut98, lam04}.

In this paper, we present the results of spectropolarimetry of the
mini-BAL quasar, HS~1603+3820. Using our data, we search for
variations in the degree of polarization across the mini-BAL trough.
The detection of a strong polarization signal in the mini-BAL trough
would lead us to infer that (a) scattering of photons around the
absorber contributes to the partial coverage, and (b) the variability
of the mini-BAL can be caused by fluctuations in the amount of
scattered light, as depicted in Figure~1A.  Furthermore, we would also
expect this fractional polarization to increase as the coverage
fraction decreases.

In \S2, we describe the observations and data reduction and in \S3 we
describe the spectropolarimetric properties of HS~1603+3820.  In \S4,
we discuss the constraints placed by our observations on possible
scattering mechanisms in HS~1603+3820 and we evaluate the scenarios
for the variability of the mini-BAL in light of the new data.  We
adopt $\zem = 2.542$ as the systemic redshift of the quasar, which was
estimated from its narrow emission lines \citep{mis03}.

\section{Observations and Data Reduction}

The spectropolarimetric observation for HS~1603+3820 was carried out
with the Faint Object Camera And Spectrograph (FOCAS) \citep{kas02} on
the Subaru Telescope, on 2008 August 7--8 (UT).  The detector is a
mosaic of two 4k$\;\times\;$2k MIT CCDs with 15$\mu$m pixels.  All the
observations were carried out through a polarimetric unit that
consists of a rotating superachromatic half-wave plate and a quartz
Wollaston prism.  Since the circular polarization is negligible for
QSOs, we measured only linear polarization.  We used a 0\farcs8 slit,
the 600 mm$^{-1}$ VPH grism (600\_520nm), and the L600 order-sorting
filter.  This setting results in a resolving power of $R \sim 1500$.
A segment of the spectropolarimetric observation consists of exposures
at four wave-plate position angles: 0\fdg0, 45\fdg0, 22\fdg5, and
67\fdg5.  The integration time of each exposure was 900 seconds, and
the total integration time was 8 hours.  The position angle of the
entrance slit was set to be 180$^{\circ}$.  We also obtained spectra
of an unpolarized flux standard star
\citep[BD+28$^{\circ}$~4211;][]{oke90} and a strongly polarized star
(HD~204827).  Spectra of a halogen lamp and a thorium-argon lamp were
also obtained for flat fielding and wavelength calibration,
respectively. The observation log is presented in Table~1.

\begin{deluxetable*}{lccrlrcccl}
\tabletypesize{\scriptsize}
\setlength{\tabcolsep}{0.04in}
\setcounter{table}{1}
\tablecaption{Log of Observations \label{t1}}
\tablewidth{0pt}
\tablehead{
\colhead{Target}       &
\colhead{RA$^a$}       & 
\colhead{Dec$^a$}      & 
\colhead{$m_V$}        & 
\colhead{Date}         &
\colhead{Exposure}     &
\colhead{S/N$^b$}      &
\colhead{$p^c$}        &
\colhead{$\theta\;^d$} &
\colhead{Note}         \\
\colhead{}             &
\colhead{(h:m:s)}      & 
\colhead{(d:m:s)}      &
\colhead{(mag.)}       & 
\colhead{}             &
\colhead{(sec)}        & 
\colhead{(pix$^{-1}$)} &
\colhead{(\%)}         &
\colhead{(deg.)}       &
\colhead{}             \\
\colhead{(1)}          &
\colhead{(2)}          &
\colhead{(3)}          &
\colhead{(4)}          &
\colhead{(5)}          &
\colhead{(6)}          &
\colhead{(7)}          &
\colhead{(8)}          &
\colhead{(9)}          &
\colhead{(10)}         \\
}
\startdata
HS1603+3820 & 16 04 55.4 & +38 12 02 & 15.9~~ & 2008 Aug 7,8 & 28800 & 175 & 0.64$\pm$0.15 & 166 & quasar \\
BD+28 4211  & 21 51 11.1 & +28 51 52 & 10.5~~ & 2008 Aug 7,8 &  1080 &     &               &     & unpolarized star \\
HD 204827    & 21 28 57.8 & +58 44 23 &  8.0~~ & 2008 Aug 8   &    24 &     &               &     & strongly polarized star \\
\enddata
\tablenotetext{a}{J2000.0 coordinates.}
\tablenotetext{b}{Typical S/N at the continuum level.}
\tablenotetext{c}{Average percentage polarization and
  r.m.s. dispersion over the observed wavelength range. }
\tablenotetext{d}{Polarization angle in degree.}
\end{deluxetable*}

The data were reduced in a standard manner, using IRAF\footnote{IRAF
  (Image Reduction and Analysis Facility) is distributed by the
  National Optical Astronomy Observatory, which is operated by the
  Association of Universities for Research in Astronomy Inc., under
  corporative agreement with the National Science Foundation.}.  We
extracted the spectra of HS~1603+3820 and the standard stars from a
4\farcs0 aperture along the slit (i.e., 13 binned pixels).  The
instrumental polarization was corrected using the spectrum of the
unpolarized standard star.  The instrumental depolarization was not
corrected because it has been found experimentally that it is less
than a few percent of the total polarization degree.  Flux calibration
was performed using the spectrum of the unpolarized standard star.
The polarization angle was calibrated using the data of the strongly
polarized star.

The Galactic interstellar medium is also responsible for partial
linear polarization, and both the degree and the position angle of
polarization show systematic variations with wavelength (e.g.,
\citealt{mes97}), which is most naturally explained in terms of the
presence of two or more different dust components \citep{ser75, coy74}.
We estimate the Galactic interstellar polarization toward the
direction of HS~1603+3820 to be $p_{\rm ISM}\le 0.19$\%, following the
procedure below.  The total Galactic \ion{H}{1} column density was
measured as $1.20\times 10 ^{20}$~\cmm\ by the Leiden/Argentine/Bonn
(LAB) Survey \citep{kal05}.  The Galactic neutral and molecular
hydrogen column density can be conventionally converted to a color
excess by $\langle N(\mathrm{H\,{\scriptstyle I}+H_2})/E(B-V) \rangle
= 5.8 \times 10^{21}~{\rm cm^{-2}~mag^{-1}}$ \citep{boh78}, which
gives $E(B-V) = 0.021$~mag.  Finally, we estimate the maximum
polarization degree ($p_{\rm max}$) using the empirical relation
$p_{\rm max} = 9.0\;E(B-V) = 0.19$\% \citep{ser75}.  The Galactic
interstellar polarization estimated above is considerably smaller than
the observed polarization ($p \sim 0.64$\%) and does not show any
clear spectral dependence in either the degree of polarization or
position angle as we describe in the following section.  The small
Galactic interstellar polarization toward the quasar is also confirmed
using the polarization catalog of Galactic stars \citep{hei00}.
Therefore, we did not correct for the Galactic interstellar
polarization.

\section{RESULTS}

Our primary results are illustrated in Figures~2 and 3. In Figure~2 we
plot as a function of wavelength the total flux density per unit
wavelength ($f_\lambda$, hereafter $f$), linear polarization degree
($p$), polarization position angle ($\theta$), and polarized flux ($f
\times p$).  We have omitted error bars because the S/N is high enough
even without binning.  An expanded view of these spectra around the
\ion{C}{4} mini-BAL feature (in the wavelength interval shaded in
Figure~2) is also shown in Figure~3 with 1$\sigma$ error bars
included.  In Figure~4 we show the normalized total flux spectrum in
the immediate vicinity of the \ion{C}{4} mini-BAL profile, compared to
previously published, higher-resolution spectra.

\begin{figure*}
\hbox to 7.1 truein{\vsize 6.3 truein 
 \vbox to 6.3 truein{\hsize 3.35truein
 \begin{center}
  \includegraphics[width=12cm,angle=270]{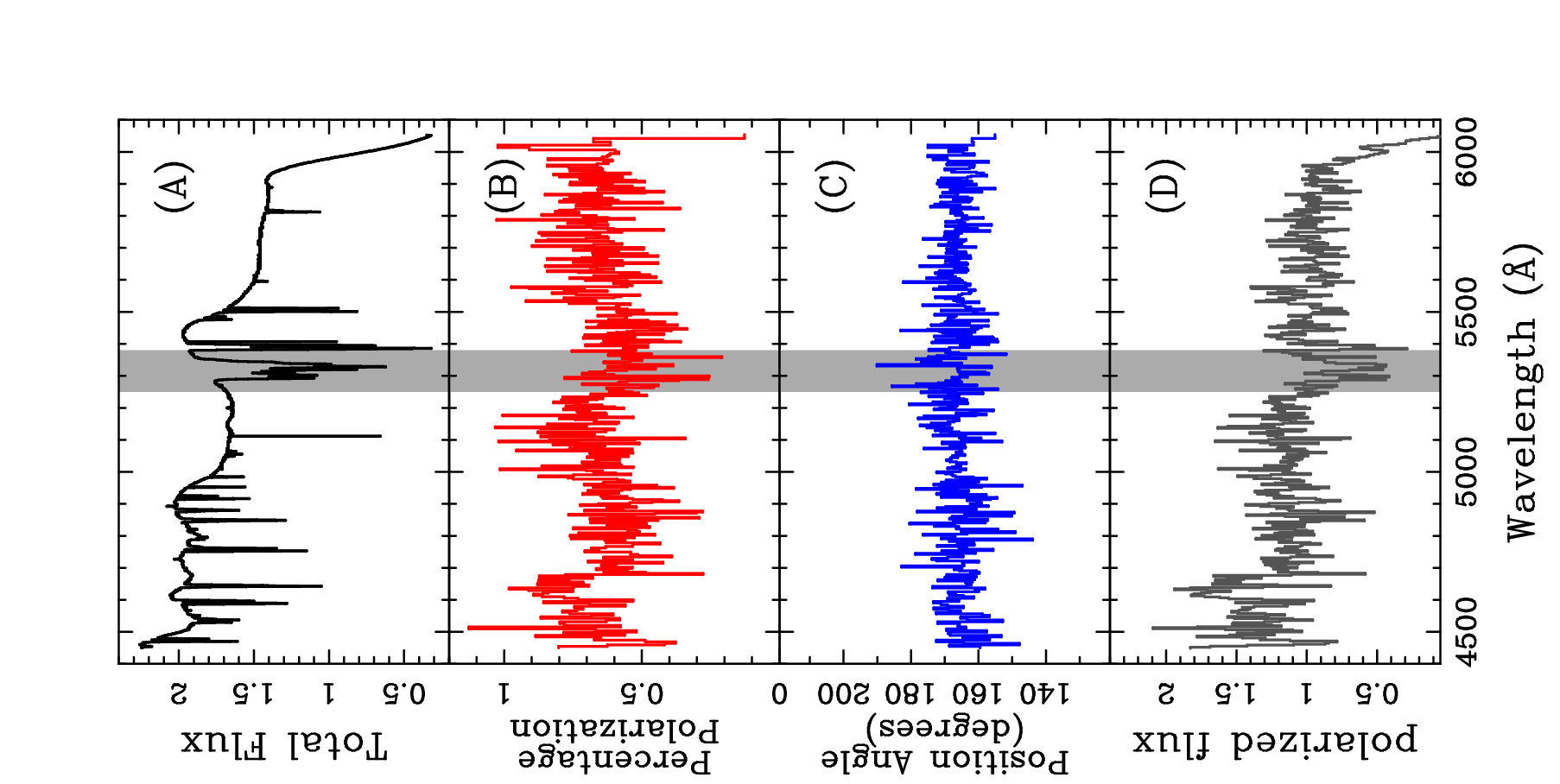}
 \end{center}
 \caption{Summary of the results of the spectropolarimetric
   observations of HS~1603+3820.  (A) the total flux density spectrum
   ($f_\lambda$ or $f$), (B) the percentage polarization spectrum
   ($p$), (C) the polarization angle ($\theta$, in degrees), and (D)
   the polarized flux (the product of the fractional polarization and
   total light spectrum, $f \times p$).  The fluxes are in units of
   $10^{-15}$ ergs~cm$^{-2}$~s$^{-1}$~\AA$^{-1}$.  The wavelength
   scale is in the frame of the observer. The shaded band shows the
   spectral region immediately around the \ion{C}{4} mini-BAL.}
 \vfill}
\hfill
 \vbox to 6.3 truein{\hsize 3.35truein
 \begin{center}
 \vskip -0.15 truein
  \includegraphics[width=12cm,angle=270]{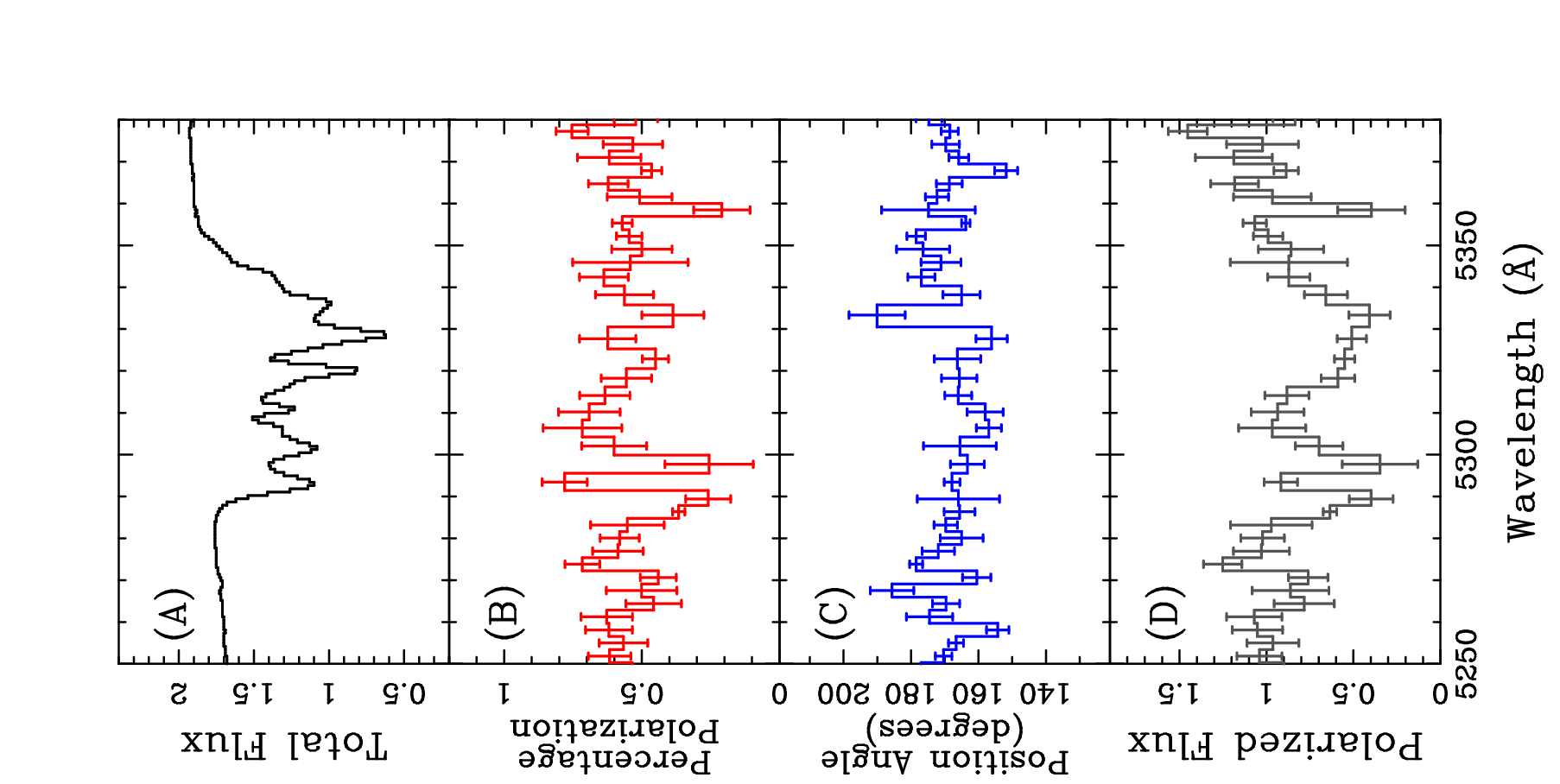}
 \end{center}
 \caption{An expanded view of the spectra of Figure~2 in the vicinity
   of the \ion{C}{4} mini-BAL. The bin size is $\sim 0.8$~\AA\ in (A)
   and $\sim 5$~\AA\ in all other panels. In panels (B)--(D) we also
   plot ``1-$\sigma$'' error bars.}
 \vfill}
}
\end{figure*}

We obtain an average degree of linear polarization of $\langle
p\rangle \approx 0.64$\% with a dispersion $\sigma_p \approx 0.15$\%
in the continuum, which is very similar to the average value in
normal, non-BAL quasars. The polarization position angle is
$\theta\approx 165$\arcdeg\ over the entire range of the observed
spectrum.  The polarization level does {\it not} rise in the
\ion{C}{4} mini-BAL trough, while it usually does in BAL profiles.  We
do not see any rotation of the polarization angle across the
\ion{C}{4} mini-BAL feature, although such a rotation is observed in
some BAL~QSOs \citep[e.g.,][]{lam04}. With the combination of high
polarization sensitivity ($\delta p \sim 0.1$\%) and high resolution
of our spectra we would have detected any change in polarization
degree or position angle rotation across the \ion{C}{4} mini-BAL
trough of HS~1603+3820, even if this is where considerably smaller
than what is typically observed in BAL quasars.

\begin{figure}
 \begin{center}
  \includegraphics[width=8cm,angle=0]{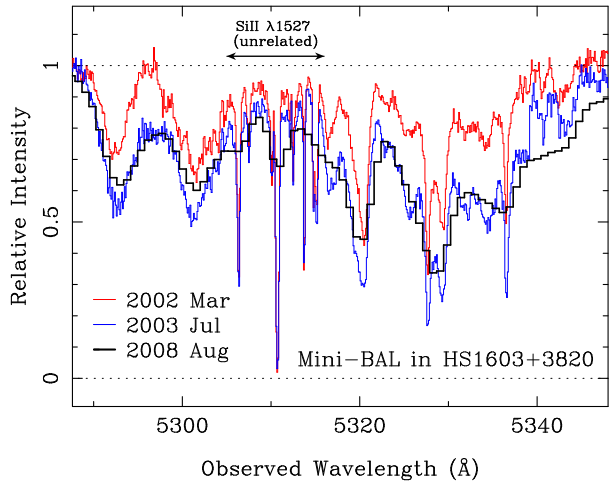}
 \end{center}
 \caption{Normalized spectra of the quasar HS~1603+3820 around the
   \ion{C}{4} mini-BAL for three epochs. The two higher-resolution
   spectra were taken with Subaru/HDS (Misawa et al. 2007b) and
   illustrate the full range of observed variability (the upper trace,
   in red, shows the spectrum from 2002 March, while the lower trace,
   in blue, shows the the spectrum from 2003 July). The coarsely
   binned, thick black line shows the lower-resolution Subaru/FOCAS
   spectrum discussed in this paper.  (obtained in 2008 August). The
   narrower \ion{Si}{2} lines are unrelated to the quasar, and arising
   at a foreground galaxy.  }
\end{figure}

We also measure the following quantities from the spectrum: a
detachment index of ${\rm DI} \approx 2.01$, a rest equivalent width
of the red part of the \ion{C}{4} broad emission line of
\ion{C}{4}$_{\rm HREW} \approx 5.5$~\AA, a logarithmic slope of the
optical spectrum of $\alpha_o \approx 0.78$, a maximum offset velocity
of the mini-BAL trough of $\Delta v_{\rm max} \approx
11,220$~\kms\ (at the left edge of the bluest absorption component at
$v_{off} \approx 10,600$~\kms), a spectropolarimetric index of ${\rm
  SI} \approx 1.18$, and a maximum polarization in the \ion{C}{4}
mini-BAL trough of $p_{\rm max} \approx 0.82$\%.  A decrease in
polarization is evident across all the emission lines in Figure~2B,
with the result that the polarized light spectrum plotted in Figure~2D
shows no emission lines.  The drop in polarization is most pronounced
in the \ion{Si}{4} and \ion{C}{4} emission lines at approximately 4900
and 5400~\AA. We measure a ratio of the polarization degree in the
\ion{C}{4} emission line to that of the continuum adjacent to it of
$p_{\rm C\; IV}/p_{\rm cont} \approx 0.8$. A similar behavior has been
observed in some BAL QSOs \citep[e.g.,][]{ogl99, lam04}.

\section{DISCUSSION}

We begin this section by discussing the implications of our
observations on possible scattering mechanisms.  We then go on to
evaluate the scattering and shielding scenarios for partial coverage,
in light of the polarization signature that we observed.

\subsection{Implications of the Observations for Scattering Mechanisms}

An important result of our observations is that the degree of
polarization does not change across the trough of the \ion{C}{4}
mini-BAL in HS~1603+3820. This result has direct implications for some
of the scattering mechanisms that may operate in this object, as we
discuss below. We specifically consider: (i) Large-angle electron
scattering in a medium off the cylinder of sight and close to the
background source, (ii) electron scattering around the mini-BAL
absorber, and (iii) resonance scattering inside the absorber. We do
not discuss scenarios involving dust scattering although we do note
that they suffer from the following drawback: the grains must be close
to the continuum source so as not to scatter a significant fraction of
BELR photons, but then they will not be able to survive the intense
radiation field at this distance.

\begin{description}
\item[]{{\it (i) Large-angle electron scattering around the background
      source. --} This is a promising mechanism, if the degree of
    polarization in the trough is large ($p \sim 10$\%), as often seen
    in spectra of BAL~QSOs. A non-uniform, anisotropic ``corona''
    around the continuum source is a candidate scattering medium
    (Figure~1A). In the case we are considering here, this mechanism
    could operate only to a small degree, i.e., the fraction of light
    that is scattered must be small so that it does not lead to a
    measurable increase in the polarization within the mini-BAL
    trough.}

\item[]{{\it (ii) Electron scattering around the absorber. --} This
    mechanism is similar to the previous one, but the scattering
    medium surrounds the absorber, not the continuum source
    (Figure~1C). This corona could be the highly ionized atmosphere of
    the absorbing gas parcel. The degree of polarization depends on
    the absorber's size and distance from the continuum source, as
    well as its geometry. Thus, a filamentary absorber with an
    extended scattering corona, located very close to the source is
    disfavored because in such a case the scattering angle (denoted by
    $s$ in Figure~1C) is large and the scattered photons will have a
    large polarization, in contradiction with our observations.  On
    the other hand, if the absorber is very far from the background
    source (such that the extent of the scatering medium is much
    smaller than the distance from the background source), the
    scattering angle is inevitably small and so is the resulting
    polarization. Such a scenario remains viable in view of our
    results. In fact, the plausibility of this scenario is reinforced
    by the finding that the absorbers in some quasars are located at
    distances of several kiloparsec or more from the continuum source
    \citep{ham01, kor08, dun10}.}

\item[]{{\it (iii) Resonance scattering inside the absorber --}
  Resonance scattering of line photons within the absorber can produce
  some level of polarization. \citet{lee94} introduced this scenario,
  while \citet{lee97} predicted a significant rise in the degree of
  polarization inside the absorption troughs.  The predicted
  polarization levels in the absorption troughs are $p\sim5$--15\%
  depending on the geometry, which is in contradiction with our
  observations Thus, we consider resonance scattering unlikely.}

\end{description}

\subsection{The Cause of Mini-BAL Variability}

\subsubsection{Scattering Scenario}
Considering the results of our spectropolarimetric observation, we can
reject with some confidence the large-angle scattering scenario
(Figure~1A) as the cause of partial coverage, hence the origin of
variability of the \ion{C}{4} mini-BAL in HS~1603+3820. The primary
argument against this scenario is that the polarization level remains
low in the mini-BAL trough and similar to the level out of the trough.
The absorption depth of the \ion{C}{4} mini-BAL in the observation
presented here is clearly smaller than the absorption depth in the
2003 July observation (see Figure~4). In the context of the scattering
hypothesis, this suggests that there should be a considerable amount
of scattered light in the absorption trough, yet we did not detect any
change in polarization across the absorption trough.  Similar
arguments apply to the resonance scattering scenario.
The only scattering scenario that remains viable is one in which the
scattering medium surrounds the distant, intrinsic absorber (as
depicted in Figure~1C). In this scenario the scattering angle is small
and so is the resulting degree of polarization. The observed coverage
fraction sets an additional constraint on this scenario: the
combination of scattering optical depth and solid angle in the
scattering medium around the absorber must be such so it can redirect
the appropriate number of photons towards the observer.

\acknowledgments The research was supported by the Japan Society for
the Promotion of Science through Grant-in-Aid for Scientific Research
21740150.  TM acknowledges supports from the Special Postdoctoral
Research Program of RIKEN and from the Sumitomo Foundation (090358).
ME and JC acknowledge support from NSF grant AST-0807993. We thank the
anonymous referee for a number of comments that helped us improve the
paper.

\end{document}